\documentclass{iau}
\usepackage{graphicx}

\title[S295.~~  The LF of TP-AGB stars in the LMC and SMC] 
{The LF of TP-AGB stars in the LMC/SMC}

\author[Gustavo Bruzual et al.]{
Gustavo Bruzual$^{1}$,
St\'ephane Charlot$^{2,3}$,
Rosa Gonz\'alez L\'opezlira$^{1}$,
Sundar Srinivasan$^{4}$,
Martha L. Boyer$^{5}$, and
David Riebel$^{6}$
}

\affiliation{
$^1$Centro de Radioastronom\'\i a y Astrof\'\i sica, UNAM, Campus Morelia, M\'exico\\ email: {\tt g.bruzual@crya.unam.mx, r.gonzalez@crya.unam.mx } \\[\affilskip]
$^2$UPMC, UMR7095, Institut d'Astrophysique de Paris, F-75014, Paris, France\\
$^3$CNRS, UMR7095, Institut d'Astrophysique de Paris, F-75014, Paris, France\\ email: {\tt charlot@iap.fr } \\[\affilskip]
$^4$Academia Sinica, Institute of Astronomy and Astrophysics, PO Box 23-141, Taipei 10617, Taiwan, R. O. C., email: {\tt sundar@asiaa.sinica.edu.tw } \\[\affilskip]
$^5$STScI, 3700 San Martin Drive, Baltimore, MD 21218, USA, email: {\tt mboyer@stsci.edu } \\[\affilskip]
$^6$Department of Physics, United States Naval Academy, 572C Holloway Road, Annapolis, MD 21402, USA, email: {\tt driebel@usna.edu }
}

\pubyear{2013}
\volume{295}  
\pagerange{xx--xx}
\jname{The intriguing life of massive galaxies}
\editors{D. Thomas, A. Pasquali \& I. Ferreras, eds.}
\begin{document}

\maketitle

\begin{abstract}
We show that Montecarlo simulations of the TP-AGB stellar population in the LMC and SMC galaxies
using the CB$^{*}$ models produce LF and color distributions that are
in closer agreement with observations than those obtained with the BC03 and CB07 models.
This is a progress report of work that will be published elsewhere.
\end{abstract}

\firstsection 
\section{Introduction}

\begin{figure}[b]
\begin{center}
\includegraphics[width=\textwidth]{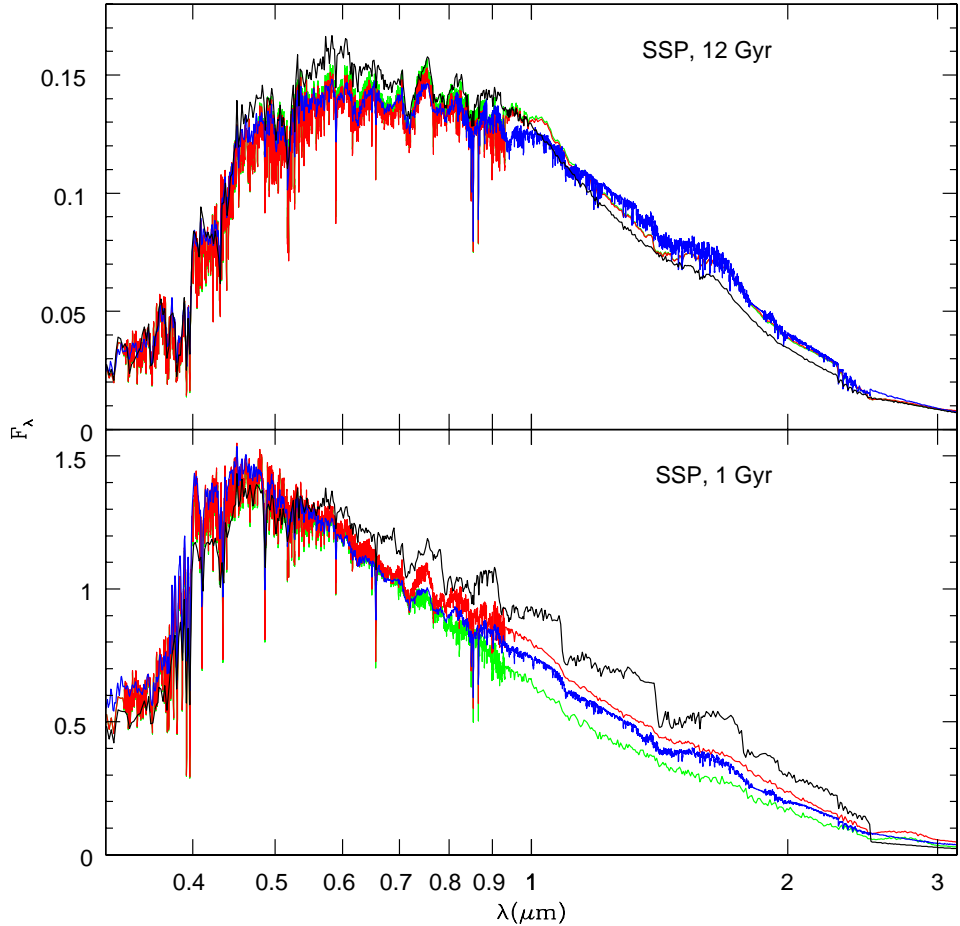}
 \caption{
SED in the optical and NIR range for a Salpeter IMF, $Z = Z_{\odot}$, SSP at the age indicated inside each frame.
The different lines clearly visible in the 1 Gyr frame in the 1-2$\mu m$ range represent from bottom to top the models by BC03,
CB$^*$ (see text), CB07, and M05. Whereas the fractional range in flux spanned by these models may reach 100\% at 1 Gyr, it is less than
15\% at 12 Gyr. \cite{mk10}, Melbourne et al. (2012), and \cite{sz12}
have shown evidence that the treatment of the TP-AGB stars in the CB07 and M05 models overestimates the contribution by TP-AGB stars in the NIR,
favoring BC03 and CB$^*$.
}
\label{fig1}
\end{center}
\end{figure}

It has been known for quite some time now that intermediate mass stars in the thermally pulsing asymptotic giant branch
(TP-AGB) phase of their evolution contribute at least 50\% of the NIR light in a simple stellar population (SSP) of
age 1-2 Gyr, e.g., \cite{cm05}, Bruzual (2007, 2011).
The treatment of this stellar phase in stellar population synthesis models determines the predicted spectral
energy distribution (SED) of stellar populations in this wavelength and age range.
In Fig.\,\ref{fig1} we compare the predictions of \cite{cm05} for a Salpeter IMF, $Z = Z_{\odot}$, SSP model at
two different ages, with the predictions of three different versions of our code:
(a) the BC03 models;
(b) the minor revision of these models introduced by CB07, and 
(c) a major revision of this code and models (in preparation, hereafter CB$^*$).
The CB07 models use the same sets of stellar tracks and spectral libraries as BC03, 
except for the TP-AGB stars, for which CB07 follow the semi-empirical evolutionary prescriptions by \cite{MG07} and \cite{MG08}. 
The CB$^*$ models used in this paper are based on the stellar evolution models computed by \cite{bertl08}.
Tracks are available for metallicities Z = {0.0001, 0.0004, 0.001, 0.002, 0.004, 0.008, $Z\odot$=0.017, 0.04, and 0.07}. 
In CB$^*$ the evolution of TP-AGB stars follows a recent prescription by Marigo \& Girardi (private communication), which has been calibrated
using observations of AGB stars in the Magellanic Clouds and nearby galaxies (Girardi et al. 2010; \cite{melb12}).
In the optical range, the CB$^*$ models are available for the IndoUS (\cite{fv04}), Miles (\cite{ps06}), 
Stelib (\cite{jfl03}), and BaSeL 3.1(\cite{pw02}) spectral libraries.
The NIR spectra of TP-AGB stars in CB$^*$ are selected from the compilation by \cite{lm02}, 
the NASA Infrared Telescope Facility (IRTF) library (\cite{jr09}), and the C-star model atlas by \cite{ba09}.
The effects of mass loss and reddening in the spectra of TP-AGB stars have been included in these models as described by
\cite{ragl10}.
The treatment of the TP-AGB in the M05 models is based on the Fuel Consumption Theorem and is thus completely independent of
the prescriptions used in the BC/CB models.

\begin{figure}[b]
\begin{center}
\includegraphics[width=\textwidth]{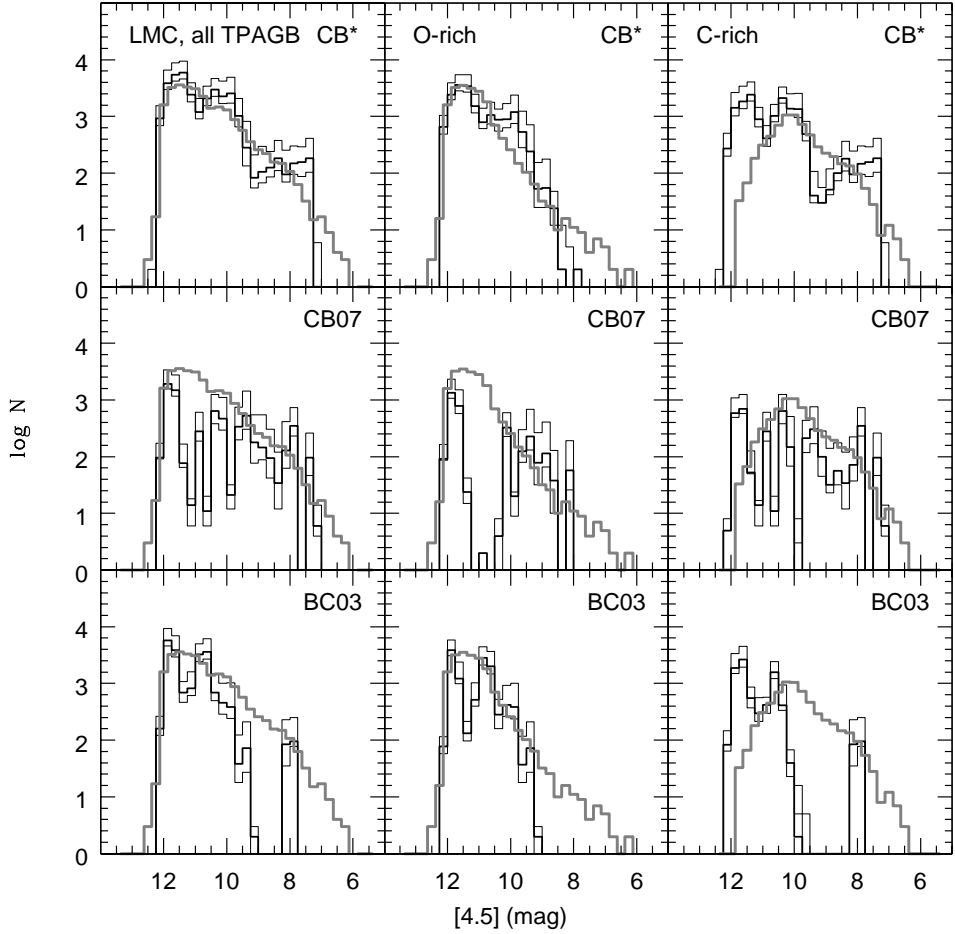}
 \caption{
LF derived from our Montecarlo simulations of the star formation history of the LMC using the
CB$^*$, CB07, and BC03 models, and from the observed $SAGE$
data set (\cite{ss09}) in the $IRAC~[4.5]\mu m$ band. The left column corresponds to all the TP-AGB
stars. In the central and right column only the O-rich and C-rich TP-AGB stars are shown, respectively.
For the simulations we assumed the Salpeter IMF, $Z = 0.008$ isochrones, and we kept all the stars with
apparent $K \le 12$ mag. The \textit{heavy gray-line} corresponds to the $SAGE$ LF. The \textit{heavy black-line}
corresponds to the simulation LF using the central value of the LMC SFH (\cite{hz09}). The bracketing \textit{light black-lines}
correspond to the upper and lower limit of the SFH derived from the error bars given by these authors.
}
\label{fig2}
\end{center}
\end{figure}

\begin{figure}[b]
\begin{center}
\includegraphics[scale=0.56]{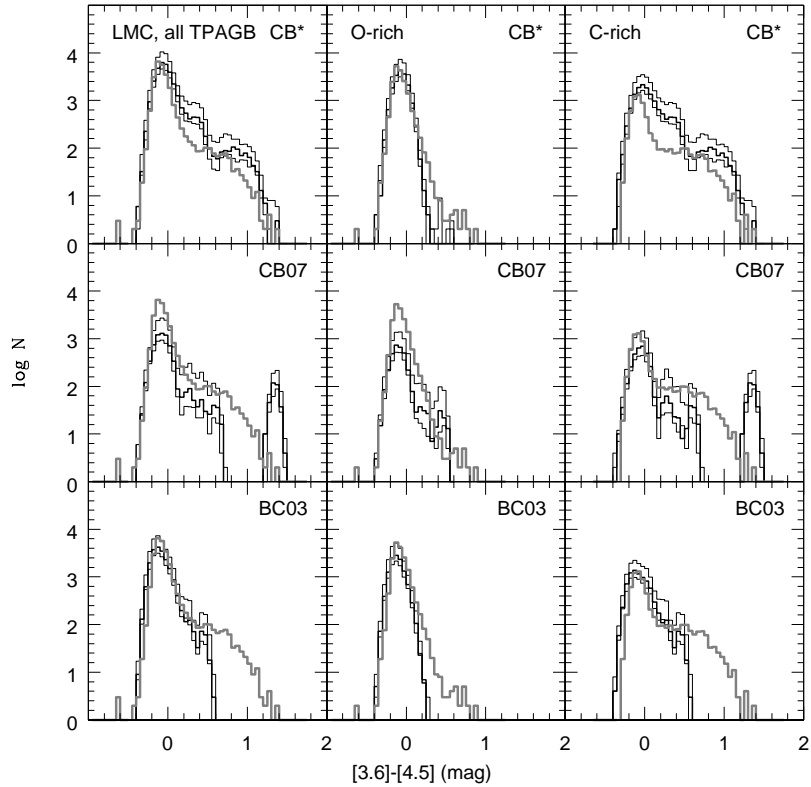}
 \caption{Same as Fig.\,\ref{fig2} but for the $IRAC~[3.6]-[4.5]\mu m$ color distribution.}
\label{fig3}
\end{center}
\end{figure}

\begin{figure}[b]
\begin{center}
\includegraphics[scale=0.56]{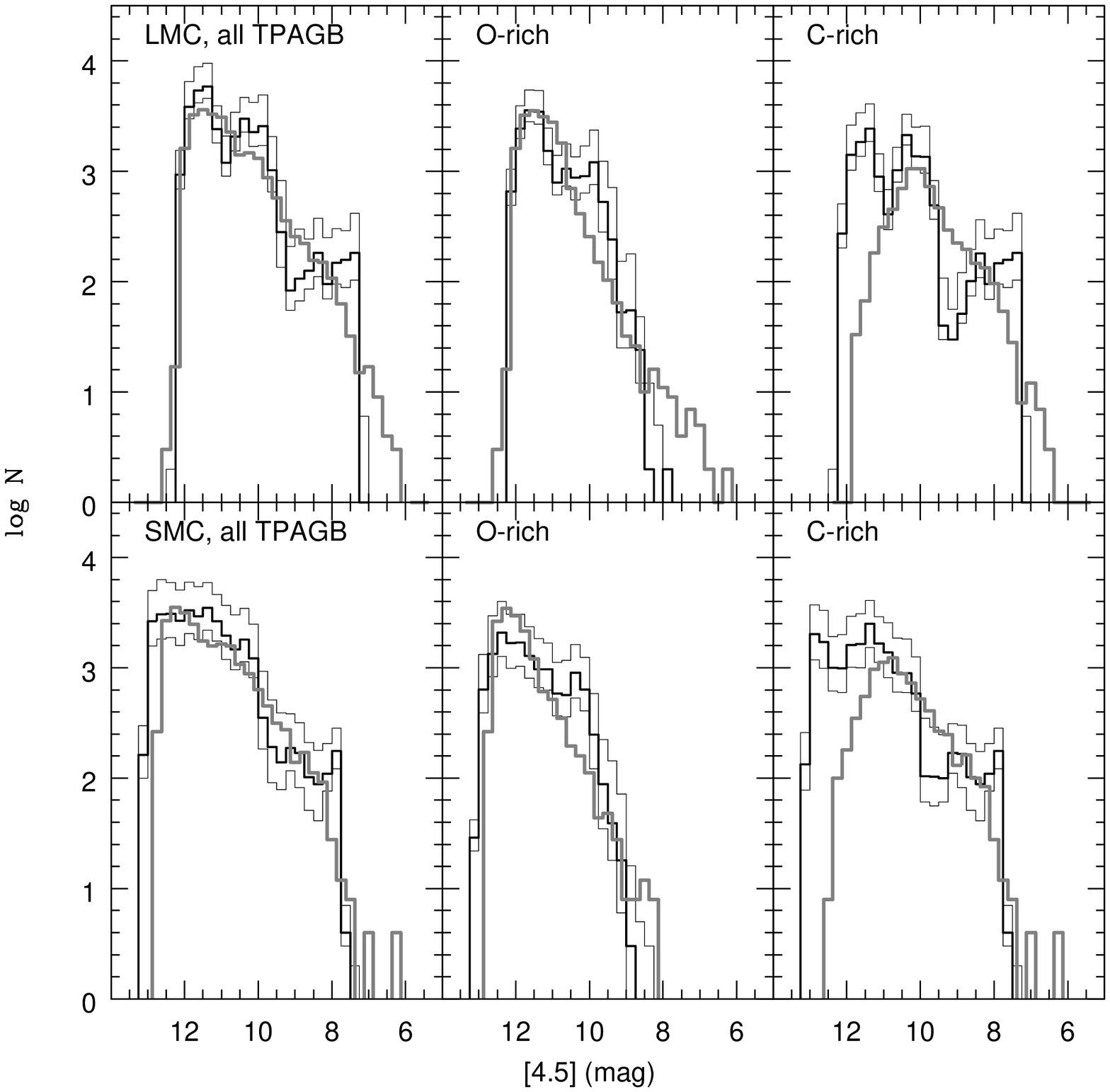}
 \caption{Same as Fig.\,\ref{fig2} but including the results of the simulations for the SMC using the SFH of \cite{hz04}.}
\label{fig4}
\end{center}
\end{figure}

\section{Modeling the LF of TP-AGB stars in the LMC and SMC}

We model the distribution of TP-AGB stars in the CMD in various optical and NIR bands for a stellar population
of $Z = 0.008$, close to the LMC metallicity, by means of Montecarlo simulations (cf. \cite{gba02}, 2010).
At each time step the mass formed in stars is derived from the LMC star formation history (\cite{hz09}).
The stars are distributed in the CMD according to the isochrones computed with the CB$^*$, CB07, and BC03
models described in \S 1.
Fig.\,\ref{fig2} shows a comparison of the LF derived from our three simulations and the observed $SAGE$
data set (\cite{ss09}) in the $IRAC~[4.5]\mu m$ band.
The corresponding $IRAC~[3.6]-[4.5]\mu m$ color distributions are shown in Fig.\,\ref{fig3}.
Using the same procedure and the SFH of the SMC from \cite{hz04} we model the TP-AGB stellar
population in the SMC galaxy, Fig.\,\ref{fig4}. In the case of the SMC the chemical evolution indicated by
\cite{hz04} is included in our simulations.
Inspection of Figs. 2-4 show that the LF's  computed with the CB$^*$ models are in closer agreement with the
observations that those computed with the BC03 and CB07 models.
These results are consistent with the findings by \cite{mk10}, Melbourne et al. (2012), and \cite{sz12} (see caption to Fig.\,\ref{fig1}),
and support our choice for the treatment of TP-AGB stars in the CB$^*$ models.
We do not have at hand enough information (isochrones) to perform the same kind of comparison with the M05 models.
Details of this work will be published in a coming paper.


G. Bruzual acknowledges support from the National Autonomous University of M\'exico through grants IA102311 and IB102212.


\begin{thebibliography}{}

\bibitem[Aringer et al. (2009)] {ba09} {Aringer, B. et al.} 2009, \textit{A\&A}, 503, 913

\bibitem[Bertelli et al. (2008)] {bertl08} {Bertelli, G. et al.} 2008,  \textit{A\&A}, 484, 815

\bibitem[Bruzual 2002] {gba02}
{Bruzual A., G.} 2002, in Extragalactic Star Clusters, IAU Symposium Ser., Vol.\ 207, eds.\
D. Geisler, E.~K. Grebel, \& D. Minniti, Provo:ASP, 616

\bibitem[Bruzual 2010] {gba10} {Bruzual, G.} 2010, \textit{RSPTA}, 368, 783

\bibitem[Bruzual (2007)]{gba07}
{Bruzual, G.} 2007, in Proceedings of the IAU Symposium No. 241 "Stellar populations as building blocks of galaxies", eds. A.
Vazdekis and R. Peletier, Cambridge: Cambridge University Press, 125 (arXiv:astro-ph 0703052)

\bibitem[Bruzual (2011)]{gba11} {Bruzual, G.} 2011, in Proceedings of the XIII LARIM, eds. W. J. Henney and S. Torres-Peimbert, \textit{Rev. Mex. Astron. Astrofis., Conf. Ser.}, 40, 36-41

\bibitem[Bruzual \& Charlot (2003)]{bc03} {Bruzual, G., \& Charlot, S.} 2003, \textit{MNRAS}, 344, 1000 (BC03)

\bibitem[Charlot \& Bruzual (2007)]{cb07} {Charlot, S., \& Bruzual, G.} 2007, unpublished, models distributed on demand (CB07)

\bibitem[Girardi et al. (2010)] {lg10} {Girardi, L. et al.} 2010, \textit{ApJ}, 724, 1030

\bibitem[Gonz\'alez-L\'opezlira et al. (2010)] {ragl10} {Gonz\'alez-L\'opezlira, R. et al.} 2010, \textit{MNRAS}, 403, 1213

\bibitem[Harris \& Zaritsky (2004)] {hz04} {Harris, J. \& Zaritsky, D.} 2004, \textit{AJ}, 127,153

\bibitem[Harris \& Zaritsky 2009] {hz09} {Harris, J. \& Zaritsky, D.} 2009, \textit{AJ}, 138,1243

\bibitem[Kriek et al. (2010)] {mk10} {Kriek M. et al.} 2010, \textit{ApJ}, 722, L64

\bibitem[Lan\c con \& Mouhcine (2002)]{lm02} {Lan\c con, A. \& Mouhcine, M.} 2002, \textit{A\&A}, 393, 167

\bibitem[Le Borgne et al. 2003]{jfl03}Le Borgne, J.-F. et al.\ 2003, \textit{A\&A}, 402, 433

\bibitem[Maraston (2005)]{cm05} {Maraston C.} 2005, \textit{MNRAS}, 362, 799 (M05)

\bibitem[Marigo \& Girardi (2007)]{MG07} {Marigo, P., \& Girardi, L.} 2007,  \textit{A\&A}, 469, 239

\bibitem[Marigo et al. (2008)] {MG08} {Marigo, P. et al.} 2008,  \textit{A\&A}, 482, 883

\bibitem[Melbourne et al. 2012] {melb12} {Melbourne, J. et al.} 2012, \textit{ApJ}, 748, 47

\bibitem[Rayner et al. 2009]{jr09}Rayner, J.T.  et al.\ 2009, \textit{ApJS}, 185, 289

\bibitem[S\'anchez-Bl\'azquez et al. 2006]{ps06}S\'anchez-Bl\'azquez, P. et al.\ 2006, \textit{MNRAS}, 371, 703

\bibitem[Srinivasan et al. 2009] {ss09} {Srinivasan, S. et al.} 2009, \textit{AJ}, 137, 4810

\bibitem[Valdes et al. 2004]{fv04}Valdes, F. et al.\ 2004, \textit{ApJS}, 152, 251

\bibitem[Westera et al. 2002]{pw02}Westera, P. et al.\ 2002, \textit{A\&A}, 381, 524

\bibitem[Zibetti et al. (2012)] {sz12} Zibetti, S. et al 2012, \textit{MNRAS}, submitted (arXiv:1205.4717)

\end{thebibliography}
\end{document}